\documentclass{appolb}
\usepackage{epsfig}

\begin{document}
 \eqsec  
\title{CONFINEMENT IN QCD: RESULTS AND OPEN PROBLEMS}
\thanks{Presented at The Cracow School of Theoretical Physics, XLV Course.}
\author{Adriano Di Giacomo}
\maketitle
\begin{abstract}
Progress is reviewed in the understanding of color confinement.
\end{abstract}
  
\section{Introduction}
\subsection{History}
 The existence of quarks was first hypothesized by M. Gell-mann in the sixties\cite{Gell1}. The history of the idea is instructive.  \\
 By analogy to the electromagnetic interaction it had been realized that the vector current of ${\beta}$ decay was the Noether current of isospin conservation (Conserved Vector Current Hypothesis\cite{FG}).\\ However the electric charge $Q$  is not a generator of the isospin symmetry group, since it contains an isoscalar part  proportional to the hypercharge  $Y=N + S$, the sum of the strangeness $S$ and of the baryon number $N$.
  \begin{equation}
     Q= T_3 + {Y\over2}
 \end{equation}
 To put weak and electromagnetic interactions on the same footing, the symmetry group had to be enlarged, to include $Y$ among the generators, to a group of rank $2$ containing the $SU(2) $
 group of isospin as a subgroup. Among  the two possible candidates $G_2$ and $SU(3)$ the latter was found to be the correct choice, with hadrons assigned to the representations $1$,$8$,$10$,$\overline{10}$, the so called eightfold way\cite{Gell2}. \\ All the representations of $SU(3)$ are sum of products of the fundamental representations $3$ and $\overline3$, so that an obvious  question  was about the existence of particles in these representations, the quarks and the antiquarks, as fundamental constituents of hadronic matter.
 Their charges as predicted by eq(1) are fractional $\pm 1/3$, $\pm 2/3$, a clear experimental signature. \\
 An intense search for quarks was immediately started, but after 40 years no quark has ever been found, and only upper limits have been established for their production cross sections and abundance.\\
 It was also realized that there was a problem with Pauli principle. If the ${\Delta}_{3/2}$ is made of three quarks, the state with charge $2$ and spin component  ${3\over 2}$ is symmetric under exchange of the quarks since for any reasonable potential the three $u$ quarks are in $S$ state. A possible way out was to assign an extra quantum number to the quarks\cite{Gell3}, which was named color, so that  each quark  could exist in three different color states.\\
 After the quantization of the gauge theories it was suggested that the color symmetry could be an $SU(3)$  gauge symmetry with quarks in the fundamental representation, and  eight  gauge bosons, the gluons mediating their interaction. The theory was named Quantum Chromodynamics (QCD)\cite{FGL}.
  Experiments provide evidence for the existence of quarks and gluons at short distances, but quarks never appear at large distances as free particles.  \\This phenomenon is known as Confinement of Color. 
  \subsection{Experiments}
  The ratio  $R\equiv{{n_q}\over {n_p}}$  of the abundances of quarks and antiquarks  to the abundance of nucleons has been investigated typically by Millikan-like experiments. No particle with fractional charge has ever been found, with an upper limit \cite{PDG}
  \begin{equation}
 R\le10^{-27}
 \end{equation}
 The expectation for $R$ in the absence of Confinement can be evaluated in the Standard Cosmological  Model\cite{Okun} as follows.\\
 At $\approx 10^{-9}$seconds after Big-Bang  when the temperature was $T\cong 10Gev$ and the effective quark mass $m_q$ of the same order of magnitude, quarks would burn to produce hadrons by the esothermic reactions   \\  $q+ \overline q \to hadrons$\\   $q + q  \to  \overline q  +  hadrons$\\
  Putting  ${\sigma_0} = \lim_{v\to0}v{\sigma}$, the burning rate  is  given by  $n_q{\sigma_0}$.
 The expansion rate in the model is  equal to  $G_N^{1/2}T^2$  with $G_N$ Newton gravitational constant ant  $T$ the temperature.  The decoupling of relic quarks will occur when due to the burning processes the quark density will decrease to a value such that the burning rate is smaller than the expansion rate, or when
 \begin{equation}
 n_q {\sigma}_0   =   G_N^{1/2}T^2
 \end{equation}
 Since the abundance of photons is  $n_{\gamma}\simeq T^3$, dividing both sides of eq(3) by $T^3$
 gives
 \begin{equation}
{n_q \over n_{\gamma}} = {G_N^{1/2}\over {T{\sigma}_0}}
  \end{equation}
 By use of the experimental values   ${n_{\gamma}\over n_p}\simeq10^9$, $G_N = 10^{-19}/m_p$, and assuming ${\sigma}_0\simeq m_{\pi}^{-2} $, $T\simeq 10Gev$ we get 
 \begin{equation}
R_{expected }\equiv  {n_q\over n_p }= {n_q\over n_{\gamma} }{n_{\gamma}\over n_p } \simeq 10^{-12}
 \end{equation}
  Quarks have been also searched as products of particle reactions \cite{PDG}, again with no result.
  As an example for the inclusive cross section ${\sigma}_q\equiv {\sigma}( p + p\to q(\overline q) + X)$ the experimental upper limit is
  \begin{equation}
{\sigma}_q  \le 10^{-40} cm^2
  \end{equation}
  The expected value in the absence of confinement  is  ${\sigma_q}_{expected}\cong {\sigma}_{Total}\cong 10^{-25} cm^2$ 
  The ratios of the upper limits to the expectations are then 
  \begin{equation}
   {R\over {R_{expected}}}\le 10^{-15}  , \\
   {{\sigma}_q\over {\sigma_q}_{expected}}\le 10^{-15}
   \end{equation}
 $10^{-15}$  is a small number. The only natural possibility is that the ratios are zero, i.e. that confinement is an absolute property, due to some symmetry of the system.   \\
 This is similar to what happens in superconductivity, where the  explanation for the upper limits on the resistivity  is that it is exactly zero, due to the Higgs breaking of the conservation of electric charge,
 or in electrodynamics where the natural explanation for the upper limit to the photon mass is that it is exactly zero, the symmetry being gauge invariance.\\
  No experimental evidence exists for the confinement of the gluons.\\
  We shall, anyhow,  define confinement as absence of colored particles in asymptotic states.
  Only  color singlet particles can propagate as free particles.\\
  As a working hypothesis we shall assume that some symmetry of the ground state is responsible for confinement.

\section{The deconfinement transition.}
 A limiting temperature exists in hadron physics, known as Hagedorn temperature \cite{Hag} $T_H$, due to the property of strong interactions to convert  excess of  energy into creation of particles. It was first conjectured in 1975\cite{CP} that its existence could be the indication of a deconfining phase transition from hadron to a plasma of quarks and gluons.\\
 This transition has not yet been detected experimentally, but extensive experimental programs and dedicated machines are being devoted to it at CERN SPS,at  Brookhaven (RHIC),and at  CERN LHC.
 The transition has been observed in Lattice QCD.\\
 Both in experiments and in lattice simulations the main problem is to define and to detect the transition, i.e. to give 
 an operational definition of confined and deconfined. In a way this problem will be the main object of my lectures.\\
 \subsection{Finite temperature QCD}
  To deal with a  system of fields at non zero temperature T one has to compute the partition function 
  \begin{equation}
   Z= Tr [ exp(-{H\over T})]
   \end {equation}
   with $H$ the Hamiltonian.\\
   It can easily be proved that  $Z$ is equal to the Feynman Euclidean path integral with the time axis compactified to the interval $(0, {1\over T})$, with periodic boundary conditions for boson fields, antiperiodic for fermions.
   \begin{equation}
   Z =  \int  [d{\phi}] e^{-\int d^3x \int ^{1\over T} _0  L[{\phi}(\vec x, t)]dt}
   \end{equation}
    A system at $T=0$ is simulated on a lattice which is in all directions bigger than the physical correlation length.  To have a finite temperature the size in the time direction $L_t$ must 
    be such that 
    \begin {equation}
    T  =  {1\over {aL_t}}
    \end{equation}
    where $a=a(\beta,m)$  is the lattice spacing in physical units, which depends on $\beta=\equiv {2N_c\over g^2}$  and on the quark masses $m$.  The size  $L_s$ in the space directions, instead, must be larger than all physical scales.  An asymmetric lattice is therefore needed   $L_t\times L_s^3$  with $L_s\gg L_t$.  \\
    The dependence of the lattice spacing  $a(\beta, m)$  on $\beta$ is dictated by renormalization group equations.  At large enough  $\beta$'s  
    \begin{equation} 
    a\cong {1\over{\Lambda}} e^{\beta\over {2b_0}}
    \end{equation}
    with $b_0$ the coefficient of the lowest order term of the beta function, which is negative because of asymptotic freedom.  For the temperature T of eq(2.3)  we obtain
    \begin{equation} 
    T\cong {\Lambda\over {L_t}} e^{\beta\over {|2b_0|}}
    \end{equation}
    $T$ is an increasing exponential function of  $\beta$, i.e. a decreasing function of the coupling constant $g^2$.    This is a peculiar behavior :  when the coupling constant is big, and the fluctuations are large, i.e. in the disordered phase  the temperature is small. In the ordered phase, instead, where
  the coupling constant and   the fluctuations are small  the temperature is large.  In ordinary thermal systems  T plays the role of the coupling constant, low temperature corresponds to order, high temperature to disorder.  \\
  The key word to understand  what happens is  Duality.
   \subsection{Duality}
   Duality is a deep concept in statistical mechanics  which has been exported  into field theory and string theory.\\
   It was first introduced in  \cite{KW}  and then developed in \cite{KC} in the frame of the 2d Ising model  which, being solvable,  is a prototype system  for it.\\
   The Ising model in 2d is defined on a simple square lattice  by associating to each site a dichotomic field variable $\sigma = {\underline +}1$. The partition function is
   \begin{equation}
   Z[\beta, {\sigma}]= \sum exp(-\sum_{ij} {\beta} {\sigma_i}{\sigma_j})
   \end{equation} 
   The sum in the action runs on nearest neighbors and  ${\beta=1/T}$ is the inverse temperature in units of the interaction constant. The model is exactly solvable.\\
   A second order Curie phase transition takes place at  $T_c={2\over {\ln(1+\sqrt2)}}$ from an ordered ferromagnetic low temperature phase in which  $<\sigma>\neq 0$ to a disordered phase in which the magnetization vanishes. \\
   The model can be considered as a discretized field  theory in (1+1) dimensions, and the lagrangean
   can be written,  apart from an irrelevant constant,  as \\ 
   $$L={\beta} \sum_{\mu=1,2} {\Delta}_{\mu} {\sigma} {\Delta}_{\mu} {\sigma}$$ \\  
   
   with
     ${\Delta}_{\mu} {\sigma} \equiv  {\sigma} (n+\hat {\mu})- {\sigma} (n)$.   
     The equation of motion is 
      ${\Delta}^2{\sigma}=0$ and a topological conserved current exists  $j_{\mu}  = {\epsilon}_{\mu\nu}{\Delta}_{\nu}{\sigma}$.  $${\Delta}_{\mu}j_{\mu} =0 $$ because of the antisymmetry of the tensor ${\epsilon}_{\mu\nu}$. 
      The corresponding conserved charge is $$ Q= \sum_{n_1}  j_0( n_0,n_1)=\sum_{n_1}{\epsilon_{01}}{\Delta_1}{\sigma} (n_0,n_1)\\= {\sigma} (n_0,+ \infty) - {\sigma}(n_0,- \infty)  $$
   In a continuum version of the model, when the correlation length goes large compared to the lattice spacing, the value at spatial infinity being a discrete variable becomes a topological quantum number.
   Typical spacial configurations with non trivial topology are the kinks  for which $\sigma$ is negative 
   below some point  $\overline n_1$ and positive above it. An anti-kink  has opposite signs.\\
    It can be shown that the operator which creates a kink $\mu(n_0,n_1)$  is a dichotomic  variable like 
    $\sigma$ and that the partition function obeys the duality equation
    \begin{equation}
    Z[\beta, \sigma] =  Z[\beta^*,\mu]
    \end {equation}
    with $$ sinh(2\beta^*)= {1\over sinh(2\beta)}$$  and the same functional form of $Z$ on both sides of eq(2.7).\\
    The system admits two equivalent descriptions : \\
    1) a 'direct'  description in terms of the fields $\sigma$  whose vacuum expectation values are the order parameters, which is convenient in the ordered phase, i.e. in the weak coupling regime. In this description  kinks are non local objects  with non trivial topology.\\
    2) a 'dual'  description in which the topological excitations become local and the original fields non local excitations. The duality mapping eq(2.7) maps the weak coupling regime of the direct description into 
    the strong coupling regime of the dual excitations and viceversa.  The dual description is convenient in the strong coupling regime of the direct description.\\
    The 2-d ising model is self-dual, being  the form of the dual partition function the same as that of the direct description, but this is not a general fact.\\
    Other examples of duality are : the duality angles-vortices
   in the 3-d X-Y model\cite{DDPT}, the duality 
    magnetization - Weiss domains in the 3d Heisenberg model \cite{DMP}, the duality  $A_{\mu}$-monopoles in compact U(1) gauge theory \cite{FM}\cite{DP}\cite{PC}, the duality fields-monopoles in N=1 SUSY  SU(2) gauge theory, and many examples in string theory\cite{ST}.\\
    The idea is then to look for dual, topologically non trivial excitations in QCD, which we shall generically denote by $\mu$, which are ordered in the confining phase $<\mu>\neq0$, thus defining the dual symmetry.     
  \subsection{The deconfinement  transition on the Lattice}  
\begin{figure}
\begin{center}
\includegraphics*[width=4in]{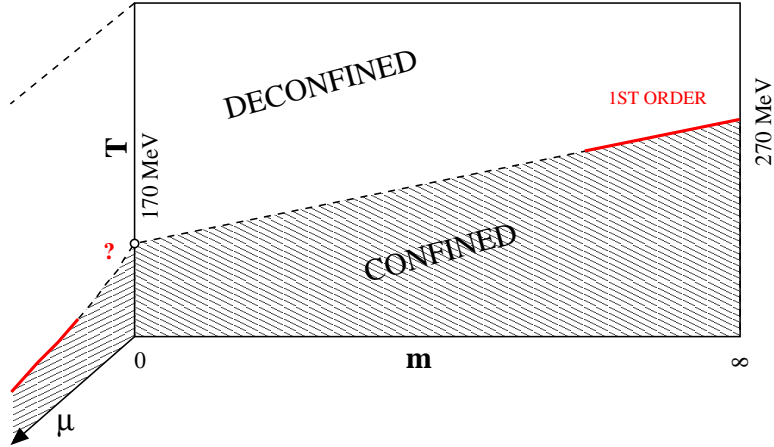}
\caption{The Phase Diagram of $N_f = 2$ QCD }
\label{1}
\end{center}
\end{figure}

  The same problem as in experiments exists for Lattice simulations :  how to define and detect the confined and the deconfined phase.\\
  In pure gauge theory (no quarks, quenched ) the Polyakov criterion is used, which consists in measuring the $q\bar q$ potential at large distances.  If it grows linearly with distance 
 
  \begin {equation}
      V(R) _{R\to\infty} \sim \sigma R
      \end{equation}
       there is confinement. If it goes to a constant 
       \begin{equation}
V(R)_{R\to\infty}  \sim   C + {C'\over R}
       \end{equation}
       the phase is deconfined.
       The potential is measured through the correlator of Polyakov lines. A Polyakov line is defined as the parallel transport along the time axis across the Lattice. 
       \begin{equation}
       L(\vec x)\equiv P exp(\int^{1\over T}_0 igA_0(\vec x,t)dt)
       \end{equation}
       In terms of the correlator of two Polyakov lines \\
       $$G(\vec x- \vec y)= <\bar L(\vec x) L(\vec y)>$$\\
       the static potential $V(\vec x- \vec y)$ acting between a quark and an antiquark is given by
       \begin{equation}
      V(\vec x- \vec y) = -T ln(G(\vec x- \vec y))
      \end{equation}
       At large distances, by cluster property,
       \begin{equation}
       <\bar L(\vec x) L(\vec y)> _{|\vec x -\vec y|\to \infty |}\approx |<L>|^2 + K exp(-{ {\sigma |\vec x -\vec     y|}\over T})
       \end{equation}
       If $|<L>|\neq 0$  then  $V(R)\to constant$ as $R\to\infty$ and there is no confinement.
       If, instead, $|<L>| = 0$ then, at large $R$,  $V(R)\approx \sigma R $ and there is confinement.\\
       $|<L>|$ is an order parameter for confinement, $Z_N$, the centre of the gauge group, being the relevant symmetry.\\
       Indeed it can be shown that 
       \begin{equation} 
       |<L>| = exp(- {F_q\over T})
       \end{equation}
       with  $F_q$ the chemical potential of a quark. In the confined phase $F_q$ diverges and   $|<L>|\to 0$.  \\
       There is  a problem in the continuum limit  since $F_q$ diverges also in the deconfined phase due to the self-energy of the quark, and a renormalization is needed\cite{ren}.\\
   \
       A transition is observed on the Lattice at a temperature $T_c$ from a low temperature phase where    $|<L>=0|$
      (confinement)  to a high temperature phase where $|<L>\neq 0|$ (deconfinement).\\
      For gauge group SU(2)  ${T_c\over \sqrt \sigma}=.50$ and the transition is second order in the universality class of the $3d$ ising model.\\
      For gauge group SU(3)   ${T_c\over \sqrt \sigma}=.630(5)$ and the transition is weak first order.
      With the usual convention  $2\pi\sigma=1 Gev$ this gives $T_c\approx 270 Mev $.
      The order of the transition is determined by use of finite size scaling techniques, which are nothing but renormalization group equations [See e.g. \cite{cardy}].\\
      The density of free energy  $F$ by dimensional arguments depends on the spacial size $L_s$ of the system in the form
      \begin{equation} 
        f =  F(L_s) L_s^4 =   \Phi ({a \over \xi}, {\xi \over L_s})
       \end{equation}
       where  $a$ is the lattice spacing and  $\xi$ is the correlation length.\\
      In the vicinity of   $T_c$      $\xi$  goes large with respect to $a$, so that  ${a\over \xi}\approx 0$.\\
      Since  $\xi$ diverges as  $\tau\equiv (1 -{T\over T_c})\to 0$ as
      \begin{equation}
       \xi    \propto \tau^{-\nu}  
      \end{equation}
      the variable ${\xi\over L_s}$ can be traded with  the variable $\tau L_s^{1\over \nu}$, and 
     $$ f = \phi (\tau L_s^{1\over \nu})$$\\
     For the specific heat $C_V = -{1\over V}{\partial^2 \over {\partial T^2}}$ and for the susceptibility
     $\chi_{<L>} \equiv  \int d^3x<\bar L(\vec x) L(\vec 0)>$ the resulting scaling laws are
     \begin{eqnarray}
     C_V-C_0 = L_s^{\alpha\over \nu} \phi_C(\tau L_s^{1\over \nu})\\
     \chi_{<L>}  = L_s^{\gamma\over \nu} \phi_{<L>}(\tau L_s^{1\over \nu})
     \end{eqnarray}
     From the measured behavior with $L_s$ of these quantities the critical indexes $\alpha, \gamma, \nu$
     can be determined, which identify the universality class of the transition.\\
     For 3d ising  $\alpha =.11, \gamma =1.43, \nu =.63$.\\
     For a weak first order  $\alpha =1, \gamma =1, \nu =1/3$.\\
         In the presence of quarks $Z_3$ is not a symmetry any more   and $<L>$ is not an order parameter. 
         The string breaks also in the confined phase and its energy is converted into pions.
         How to define confined and deconfined?\\
         The phase diagram for the case $N_f=2$ with two quarks of equal mass $m$  is shown in 
        Fig.\ref{1}.
         A line exists across which $<L>, <\bar\psi \psi>, <E> $ all experience a rapid change, so that their susceptibilities have a peak. All these peaks  happen to coincide within errors. Conventionally the phase below the line is called confined, the one above it deconfined.  An order parameter is needed, which must exist if, as we have argued, the transition is order-disorder. The dual excitations have to be identified.
     
    \section{The dual excitations  of  QCD}
    The general idea is that the low temperature phase of QCD (strong coupling) can be described, in a dual language, in terms of topologically non trivial excitations which are non local in terms of gluons and quarks, but are local fields in a dual language, and weakly coupled\cite{'tHooft78}\cite{'tHooft81}\cite{SW}.\\
    There exist  two  main proposals for these excitations, both due to 'tHooft.\\
    
    1) Vortices\cite{'tHooft78}\\
    
    2) Monopoles\cite{'tHooft81},\cite{'tHooft75},\cite{mandl}.\\
    
In the vicinity of the deconfining transition $(T\leq T_c)$ the free energy density should depend on the dual fields in a form dictated by symmetry and scale invariance. The deconfining transition is a change of symmetry : the disorder parameter  $<\Phi_{dual}>\neq 0$  for    $T <T_c$, $<\Phi_{dual}>=0$ for

    $T \ge T_c$.\\
    Two main approaches have been developed in the literature:\\
    a) Expose in the lattice configurations the dual excitations and show that by removing them confinement gets lost. (Vortex dominance, abelian dominance, monopole dominance)\\
    b) Study the symmetry and the change of symmetry across $T_c$.
    \\
    \subsection{Vortices}
    Vortices are one dimensional defects associated to closed  lines  $C$, $V(C)$.  If $W(C')$ is the Wilson loop, i.e. the parallel transport along the line $C'$ then
    \begin{equation} 
    V(C)W(C') = W(C') V(C)e^{i{{n_{CC'}\pi }\over N_c}}
    \end{equation}
    $n_{CC'}$ is the linking number of the two curves $C,C'$, which is well defined in $3d$.\\
    In $(2+1)d$,  $<V(C)>\neq 0$  means spontaneous breaking of a symmetry, the conservation of the number of vortices minus the number of anti-vortices, $<V(C)>=0$ means super-selection of that number,
    and $V(C)$ can be an order parameter for confinement. In $(3+1)d$ this statements have no special meaning.\\
    In any case, as a consequence of Eq(3.1) whenever $V(C)$ obeys the area law  $W(C')$ obeys the perimeter law, and viceversa whenever  $W(C')$ obeys the area law $V(C)$ obeys the perimeter law.\\
    The 'tHooft loop, defined as the expectation value of a vortex going straight across the lattice, or the dual of the Polyakov line, is non zero in the confined phase, zero in the deconfined phase\cite{ddl}.
    The corresponding symmetry is  $Z_3$, which, however, does not survive the introduction of dynamical quarks.\\
    \subsection{Monopoles}
    Monopoles exist as solitons in Higgs gauge theories with the Higgs in the adjoint representation \cite{t}\cite{p}. They are stable for topological reasons.\\
    If the gauge group is $SU(2)$ they are  hedgehog-like configurations for the Higgs field $\phi$, with $\phi^i (\vec r)\propto r^i$, and are characterized  by a zero of $\phi$ corresponding to the position of the monopole.  These configurations are called monopoles because of the non trivial topology of the mapping of the sphere at spacial infinity $S_2$ on the sphere of the possible values of  $<\vec \phi>$.\\
    Physically this can be understood  in terms of the 'tHooft tensor, $F_{\mu \nu}$.
    \begin{equation}
    F_{\mu \nu}\equiv  \hat{\phi}\vec G_{\mu \nu} - {1\over g}\hat{\phi}(D_{\mu}\hat{\phi}\wedge D_{\nu}\hat{\phi})
    \end{equation}
    where $g$ is the gauge coupling constant,  $D_{\mu}\hat\phi$ the covariant derivative of $\phi$,\\
    $$D_{\mu}\hat \phi= [\partial_{\mu}- g\vec A_{\mu}\wedge ]\hat \phi$$ and $\hat\phi\equiv {{\vec \phi}\over {|\vec \phi|}}.$\\
    $F_{\mu\nu} $ is gauge invariant by construction. Moreover the bilinear terms in $A_{\mu}A_{\nu}$ cancel between the two terms of eq(3,2) and\\
    $$ F_{\mu \nu}= \partial_{\mu} (\hat \phi \vec A_{\nu} ) - \partial_{\nu} (\hat \phi \vec A_{\mu})-{1\over g} \hat \phi (\partial_{\mu} \hat \phi \wedge \partial_{\nu} \hat  \phi)$$\\
    In the unitary gauge $\vec \phi = (0,0,1)$, the last term vanishes and\\
    $$F_{\mu\nu} = \partial_{\mu} A^3_{\nu} - \partial_{\nu} A^3_{\mu} $$\\
    an abelian field.\\
    $F_{\mu\nu}$ obeys Bianchi identities 
    \begin{equation}
    \partial_{\mu} F_{\mu\nu}^*=0
    \end{equation}
     with $F_{\mu\nu}^*\equiv {1\over 2} \epsilon _{\mu\nu\rho\sigma} F_{\rho\sigma}$ the dual tensor. The identity can be violated at the location of singularities, where a non zero magnetic current exists\\
    $$\partial_{\mu} F_{\mu\nu}^*\equiv  j_{\nu}$$\\
    In any case due to the antisymmetry of  $F_{\mu\nu}^*$\\
    \begin{equation}
    \partial_{\nu} J_{\nu}=0
    \end{equation}
    
    For the monopole solution \cite{t}  \\
   
   $$ F_{\mu 0}= 0 ,    (\vec E=0) $$  \\
    $${1\over 2}\epsilon_{ijk} F_{jk}= {1\over {2g}} {r_i\over {r^3}} + Dirac- string$$\\
   
    A Dirac monopole. The string is produced by the singularity of the transformation to the unitary gauge
    at the zero of $\phi$.  The transformation to the unitary gauge is called Abelian Projection.\\
    For $SU(N)$ gauge group one can inquire   about the existence of monopole solitons and what the analog of $\hat \phi$ is\cite{ddlp}.
    If we denote  by  $\Phi= \Sigma _a \Phi^a T^a $ the Higgs field, by  $A_{\mu}= \Sigma _a A^a_{\mu} T^a$ the gauge field and by $G_{\mu\nu}=\partial_{\mu} A_{\nu}-\partial _{\nu} A_{\mu} +ig [A_{\mu},A_{\nu}]$ the field strength tensor, with $T^a$ the generators of the gauge group in the fundamental representation,
    normalized  as  $Tr(T^a T^b)= \delta ^{ab}$,we can define the generalized 'tHooft tensor as
    \begin{equation}
    F_{\mu \nu}= Tr ( \Phi G_{\mu\nu} - {i\over g} \Phi [D_{\mu} \Phi, D_{\nu} \Phi])
    \end{equation}
    The necessary and sufficient condition to have abelian projection, i.e. cancellation of bilinear terms 
    in $A_{\mu}A_{\nu}$  is that   \\
    $$ \Phi = \Phi^a  ,    \Phi^a = U(x)^{\dagger }\Phi^a_{diag} U(x) $$\\
    with U(x) an arbitrary gauge transformation and
    \begin{equation}
    \Phi^a_{diag} = diag({a\over N},{a\over N},...{a\over N}^{(N-a) times}, -{{(N-a)}\over N},-{{(N-a)}\over N}....-{{(N-a)}\over N})     
    a= 1, 2,....(N-1)
    \end{equation}
    The residual symmetry is $SU(a)\otimes SU(N-a)\otimes U(1)$.\\
    For each  $\Phi^a $ one has
    \begin{equation}
    F_{\mu \nu}^a= Tr ( \Phi^a G_{\mu\nu} - {i\over g} \Phi^a [D_{\mu} \Phi^a, D_{\nu} \Phi^a])\\=
    \partial_{\mu} Tr(\Phi^a A_{\nu}) -  \partial_{\nu} Tr(\Phi^a A_{\mu})-{i\over g} Tr(\Phi^a [\partial _{\mu} \Phi^a, \partial_{\nu} \Phi^a])
    \end{equation}
    Transforming to the unitary gauge where $\Phi^a=\Phi^a_{diag}$ gives
    \begin{equation}
    F_{\mu \nu}^a= \partial_{\mu} Tr(\Phi^a_{diag} A_{\nu}) -  \partial_{\nu} Tr(\Phi^a_{diag} A_{\mu})
    \end{equation}
     Expanding  the diagonal part of $A_{\mu}$ as a sum of simple roots of the algebra of the group,
     $\alpha ^a$,  which obey the orthogonality relations  $Tr(\alpha ^a \Phi ^b_{diag})= \delta^ {ab}$,  $A^{\mu}_{diag} = \Sigma _a  A^a_{\mu} \alpha^a$,one gets
     \begin{equation}
     F^a_{\mu\nu} =  \partial_{\mu} A^a_{\nu} -  \partial_{\nu} A^a_{\mu}
     \end{equation}
     which is an abelian field.  The simple roots have the form  \\
     $$\alpha ^a = diag( 0, 0,....0,1,-1,0,... 0)$$ \\
     with the 1 at the a-th entry.
      A monopole soliton solution exists for each value of $a$ in the $SU(2)$ subspace spanned by 
      the elements $+1$ and $-1$.
       For the Higgs  field one has\\
     $$\Phi (x) = U(x)^{\dagger } \Phi (x)_{diag} U(x) $$\\
     where  $ \Phi (x)_{diag}$ is defined with eigenvalues in decreasing order.
     Expanding  $\Phi (x)_{diag}$ in the complete basis $\Phi^a_{diag}$,\\
     $$\Phi (x)_{diag} = \Sigma_a  c^a (x) \Phi^a_{diag}$$\\
     one gets 
     \begin{equation}
     \Phi(x) = \Sigma_a c^a(x) \Phi^a(x)
     \end{equation}
     The transformation $U(x)$ is singular at the sites where some  $c^a(x)$ vanishes, i.e. wherever 
     two subsequent eigenvalues of $\Phi$ coincide: these points are the locations of the monopoles.
     The field strength $F^a_{\mu\nu}$ can be defined also in the absence of a Higgs field in the lagrangean simply as
     \begin{eqnarray}
    F_{\mu \nu}^a= Tr ( \Phi^a G_{\mu\nu} - {i\over g} \Phi^a [D_{\mu} \Phi^a, D_{\nu} \Phi^a])\\
    {\Phi}^a(x) =U^{\dagger} (x) \Phi^a_{diag} U(x)
     \end{eqnarray}
    U(x)  an arbitrary gauge transformation which can have non trivial topology or singular points.
    
    $F^a_{\mu\nu}$ depends on the choice of $U(x)$. It obeys the Bianchi identities Eq.(3.3), apart from singularities where the magnetic current can be non zero.\\  
    In any case the magnetic current will obey 
    the conservation law eq(3.4).\\
    The theory has (N-1) topological symmetries built in, corresponding to the conservation of magnetic charges.\\
    If these symmetries are realized a la Wigner the Hilbert space will be superselected. If they are Higgs broken the system will be a dual superconductor.\\
    Our working hypothesis will be that the dual symmetry of QCD is the conservation of (N-1) magnetic charges.The change of symmetry at  $T_c$ is a transition from Higgs-broken to superselected.
    Dual excitations carry magnetic charge.\\
    Our program will then be to construct  magnetically charged operators  $\mu^a$ and study their
    vacuum expectation values  $<\mu^a>$.\\
    $<\mu^a>\neq 0$      means dual superconductivity.\\
    $<\mu^a>=0$      means normal vacuum.\\
    This should hold both in quenched theory and with dynamical quarks, in agreement with the ideas of
    $N_c\to \infty$ limit of QCD.
    \subsection{Construction of  $<\mu^a>$}
    The basic idea is simply that
    \begin{equation}
    e^{ipa}|x> = |x + a>
    \end{equation}
    if $x$ is a position variable and $p$ its conjugate momentum.\\
    Specifically
    \begin{equation}
    \mu^a(\vec x, t) = exp ( i\int d^3y Tr[\Phi^a(\vec y, t) \vec E (\vec y, t)]\vec b_{\perp} (\vec x -\vec y))
    \end{equation}
    where $\vec E$ is the electric field operator, and $\vec b(\vec x-\vec y)_{\perp}$ is the vector potential produced by a static monopole sitting at $\vec x$  in $\vec y$.\\
    $$\vec \nabla \vec b_{\perp}=0,  $$ 
   $$        \vec \nabla\wedge \vec b_{\perp }={2\pi\over g}{\vec r\over r^3} + Dirac- string $$\\
   $$\Phi^a(x)= U^{\dagger}(x)\Phi^a_{diag} U(x)$$ with U(x)  a generic gauge transformation.\\
   $\mu^a$ is gauge invariant. In the gauge  $$\Phi^a=\Phi^a_{diag}$$
   \begin{equation}
   \mu^a(\vec x,t) =exp(i\int Tr[\Phi^a_{diag}\vec E(\vec y, t)]\vec b_{\perp} (\vec x-\vec y)d^3y\\
                              =exp(i \int  \vec E^a_{\perp}(\vec y,t)\vec b_{\perp}(\vec x-\vec y)d^3y
   \end{equation}
   $\vec E^a_{\perp}(\vec y,t)$ is the conjugate momentum to $\vec A^a_{\perp}(\vec y, t)$ so that
   \begin{equation}
   \mu^a(\vec x,t)|\vec A^a_{\perp} (\vec y, t)>  =   |\vec A^a_{\perp} (\vec y, t) + \vec b_{\perp} (\vec x -\vec y)>
   \end{equation}
   A Dirac monopole has been added to the abelian projected configuration.  There are (N-1) species of monopoles, corresponding to  a=1,.....  N-1.\\
   $\mu^a$ creates a singularity (monopole) in a selected gauge and in all the gauges obtained from it by a transformation which is continuous in a neighborhood  of the singularity. The number of monopoles per $fm^3$
   is finite as illustrated in  Fig.2 where an histogram is displayed of the distribution of the difference between two eigenvalues of   a plaquette operator  for different values of the lattice spacing\cite{ddlpp}.
   Therefore creating a monopole in an abelian projection implies that a monopole is also created in any other abelian projection, apart from a set of zero measure.  The statements $<\mu^a>=0$ and $<\mu^a\neq0>$ are independent of the abelian projection, so that the statement that QCD vacuum is or is not a dual superconductor are absolute, projection independent statements.\\
   \subsection{Measuring   $<\mu^a>$}
   By construction \begin{figure}
\begin{center}
\includegraphics*[width=4in]{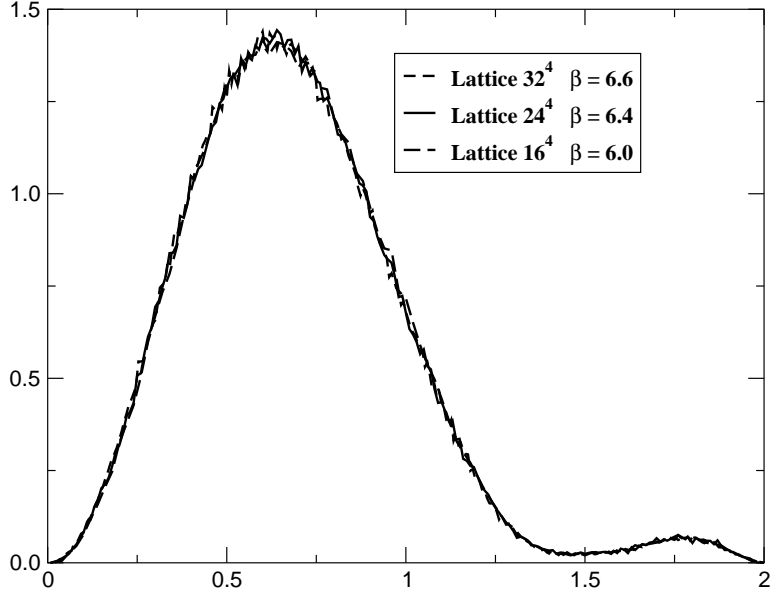}
\caption{Distribution of the differences of the phases of the eigenvalues of the Polyakov line, for three lattices with the same physical volume and different lattice spacing. A monopole on any site would correspond  to a non  zero value at zero angle. }
\label{2}
\end{center}
\end{figure}
   \begin{equation}
   <\mu^a>  ={Z^a\over Z}
   \end{equation}
   where  $Z$  is the partition function of the theory, and $Z^a$ the one modified by the insertion of the 
   monopoles. Eq(3.17) implies that  at $\beta=0$  $<\mu^a>=1$.\\
   Taking advantage of that it is convenient, instead of measuring $<\mu^a>$ directly, to measure its susceptibility  $\rho^a= {\partial\over \partial \beta}ln<\mu^a>$ which is much less noisy and will prove more suitable for our purposes. From Eq(3.17) one immediately gets 
   \begin{equation}
   \rho^a = <S>_S  -  <S^a>_{S^a}
   \end{equation}
   One also has
   \begin{equation}
   <\mu^a> = exp[\int d\beta'\rho^a(\beta')]
   \end{equation}
   It follows from Eq(3.19) that, in the infinite volume limit :\\
  (i)   $<\mu^a>\neq 0$  for $T,T_c$  iff  $\rho^a$ tends to a finite limit.\\
  (ii)  $<\mu^a>=0$   for   $T>T_c$   iff    $\rho^a\to \infty$
  The property  (ii) is much easier to check on $\rho$ than by a direct measurement of $<\mu>$ which can only give limits, due to statistical errors.\\
  In the critical region $T\approx T_c$ a strong negative peak is expected due to a rapid decrease of $<\mu^a>$, and scaling laws corresponding to the fact that the correlation length goes large with respect to the lattice spacing.
  The renormalization group equations read\cite{ddp}
  \begin{equation}
  <\mu^a> =  L_s^\kappa \Phi^a(\tau L_s^{1\over \nu},mL_s^{y_h})
  \end{equation}
  Sending $L_s\to\infty$ keeping    $\tau L_s^{1\over \nu}$ fixed gives \cite{ddp}
  $$<\mu^a> \approx m^{{-\kappa}\over {y_h}}\phi^a (\tau L_s^{1\over \nu})$$
  or  $$\rho^a /{L_s^{1\over \nu}}= f(\tau L_s^{1\over \nu})$$
  a scaling law from which the critical index $\nu$ can be determined.\\
  The prototype theory is compact $U(1)$ in $4d$, where everything is understood analytically
  at the level of theorems \cite{FM} \cite {DP}\cite{DGP}.  There is a phase transition at $\beta_c \approx 1.01$ which is first order, from a confined phase to a deconfined phase, and $<\mu>$ is non zero below
  $\beta_c$ and zero above $\beta_c$.\\Moreover $\mu$ is proved to be a gauge invariant charged operator of the Dirac type.
   A numerical determination provides a check of the approach. The result is shown in Fig.s \ref {3},\ref{4},\ref{5}
    \begin{figure}
\begin{center}
\includegraphics*[width=4in]{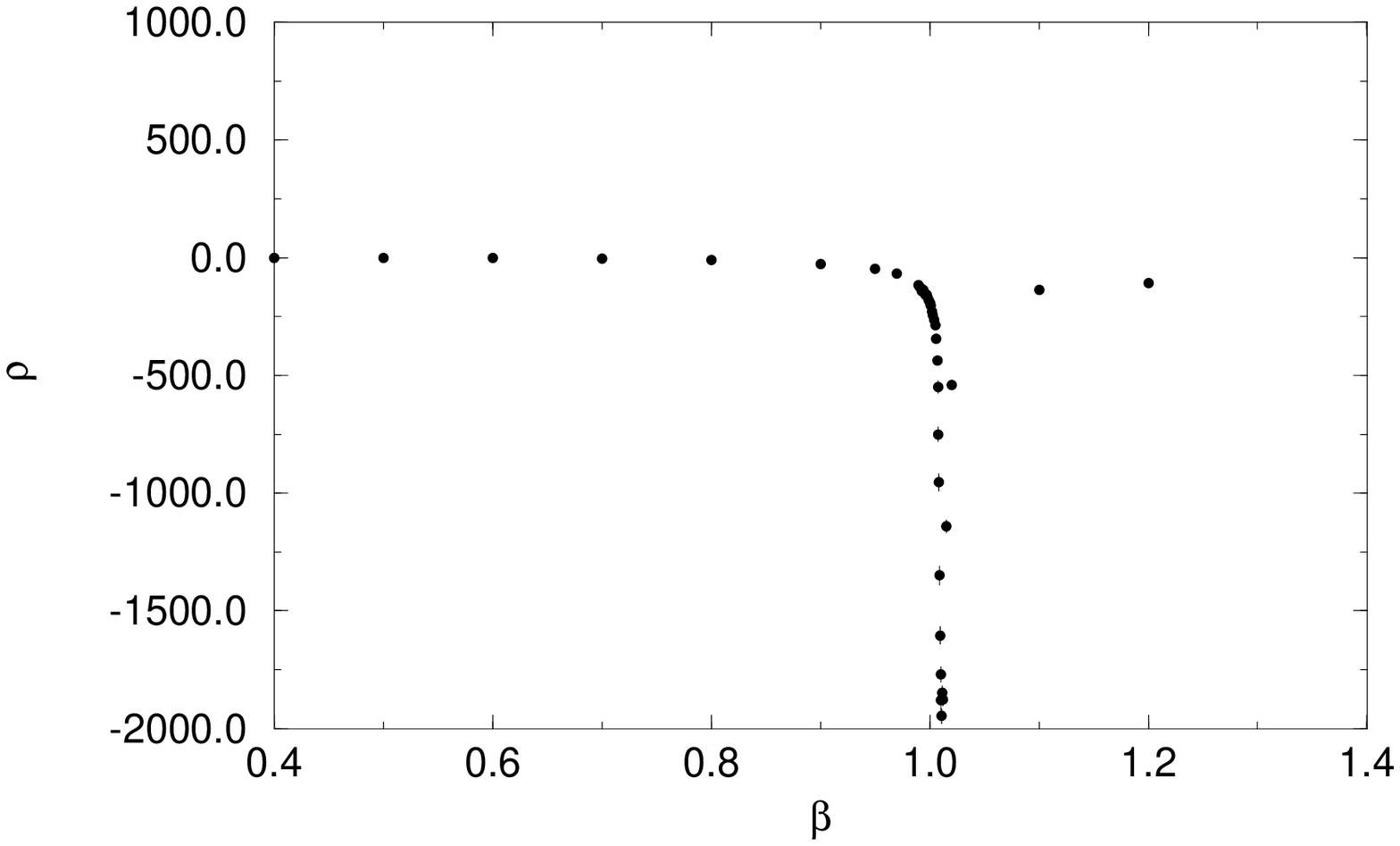}
\caption{$\rho$ vs $\beta$. The peak signals the transition }
\label{3}
\end{center}
\end{figure}

   A strong negative peak signals the transition. At low $\beta 's$ $\rho$ is size independent, at large
   $\beta 's$  it is proportional to $L_s$ with a negative coefficient, implying that $<\mu>$ is strictly zero in the thermodynamic limit.
   \begin{figure}
\begin{center}
\includegraphics*[width=4in]{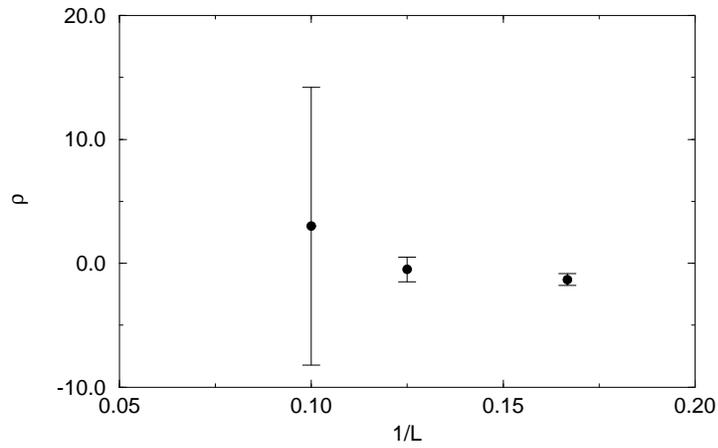}
\caption{Size dependence of $\rho$ below $\beta_c$ }
\label{4}
\end{center}
\end{figure}
 
Finite size scaling agrees with a first order transition.
 \begin{figure}
\begin{center}
\includegraphics*[width=4in]{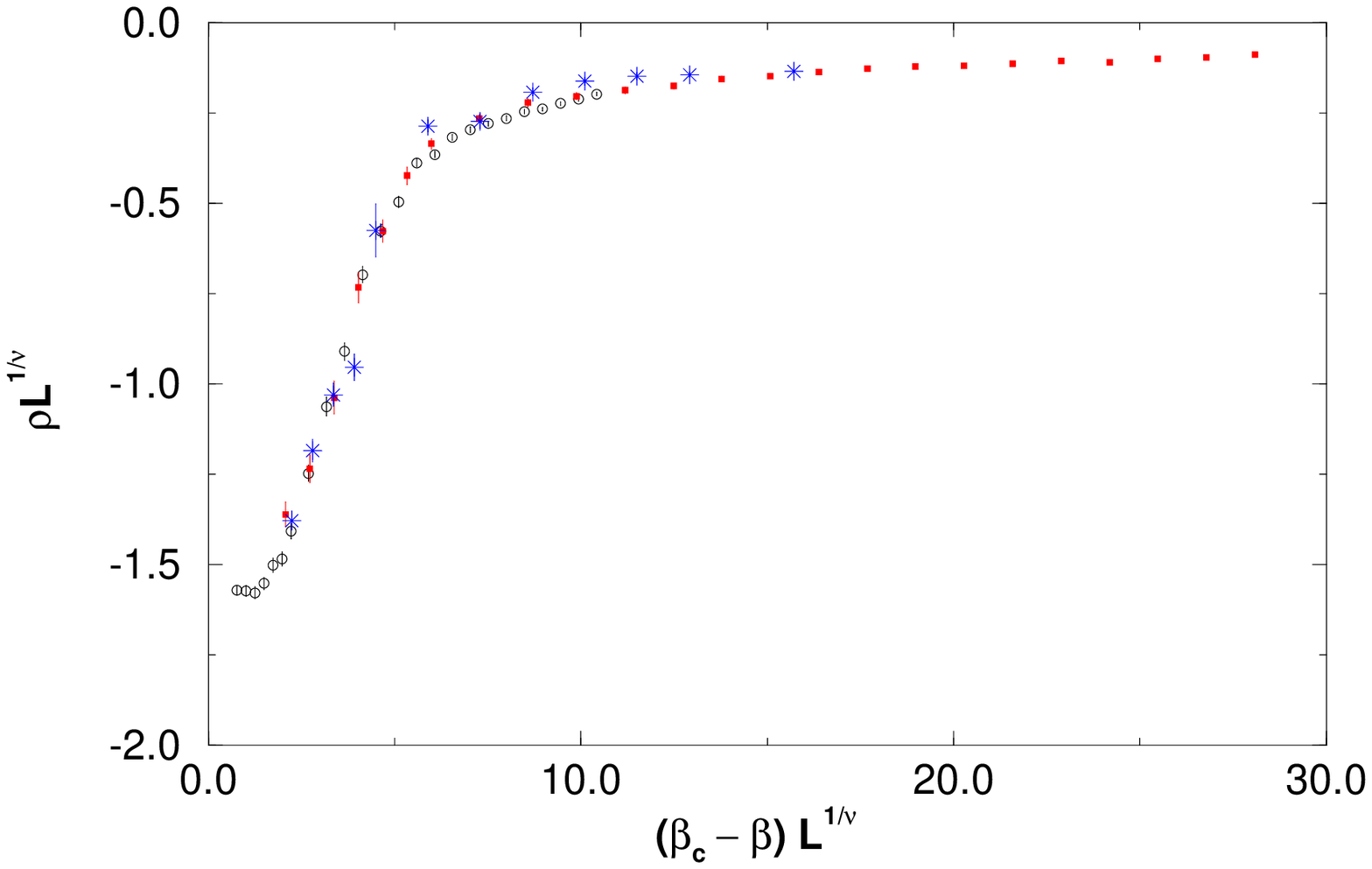}
\caption{Finite size scaling. }
\label{5}
\end{center}
\end{figure}
   
For quenched $SU(2)$ theory the deconfining transition is detected in a similar way at the right value of $\beta$ and the critical index $\nu$ is that of the $3d$ ising model\cite{dlmp1}.

   Quenched $SU(3)$ also shows a first order transition at the right tem\-perature \cite{dlmp2}.

   The numerical check of the independence on the choice of the abelian projection is contained in \cite{cddlp}.

\begin{figure}
\begin{center}
\includegraphics*[width=4in]{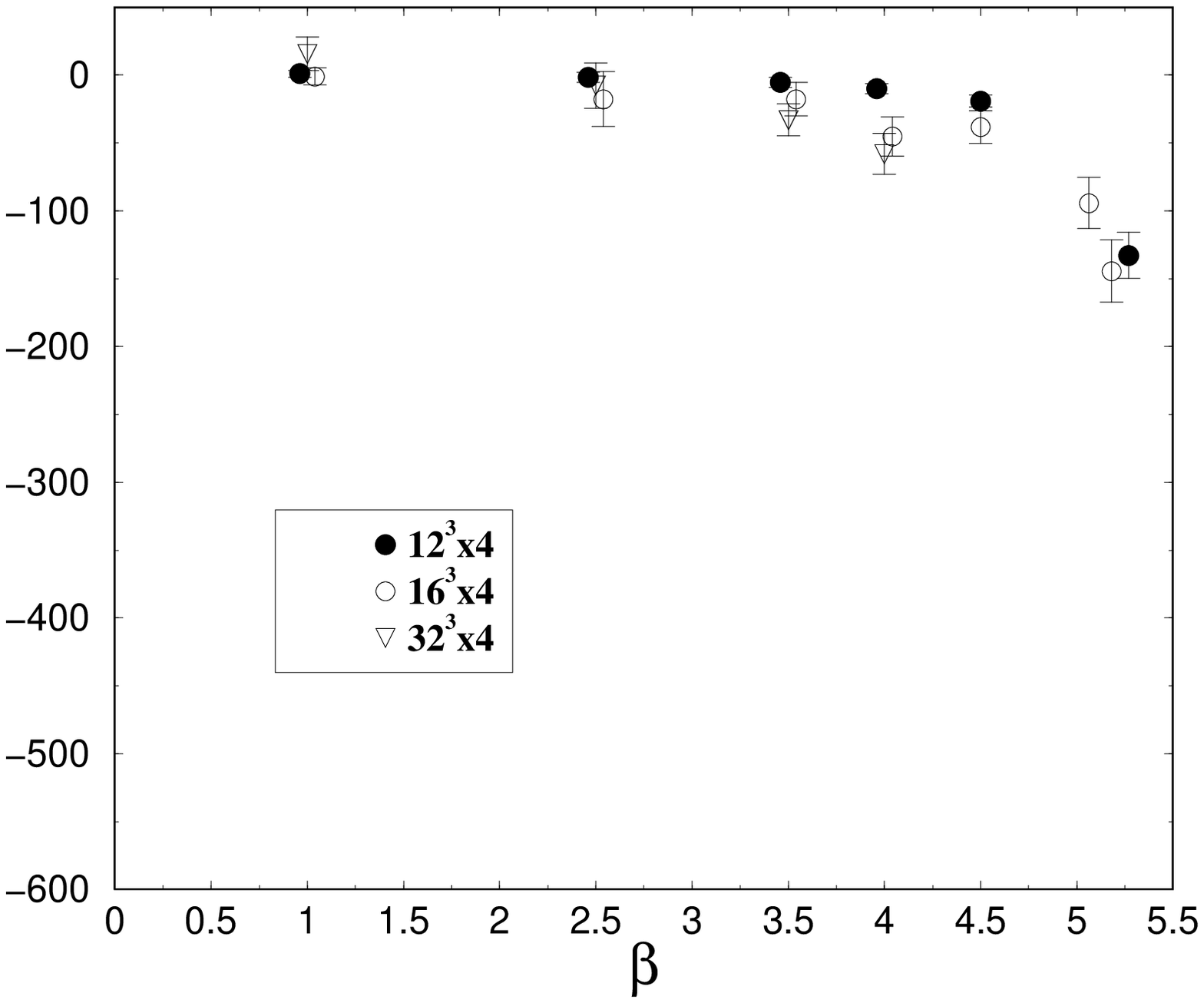}
\caption{$N_f=2$.Size dependence of $\rho$ below $\beta_c$ }
\label{6}
\end{center}
\end{figure}

\begin{figure}
\begin{center}
\includegraphics*[width=4in]{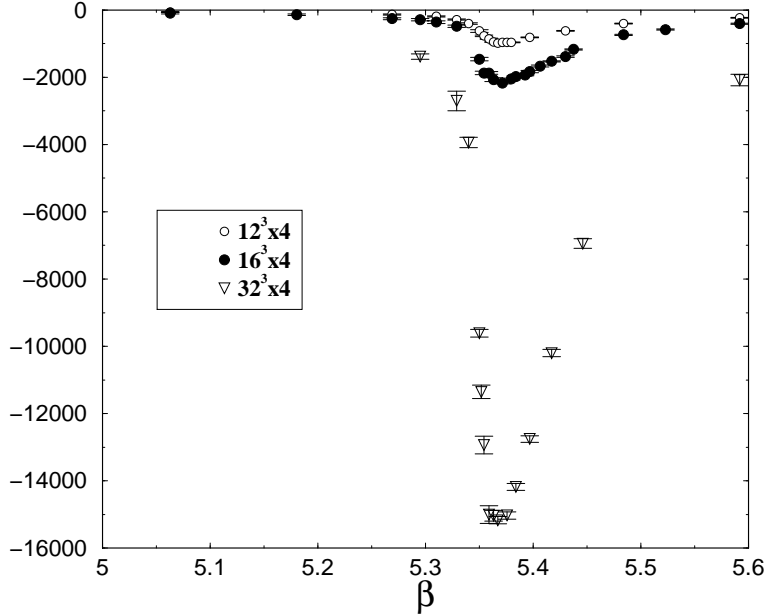}
\caption{$\rho$ peaks at different spatial sizes. }
\label{7}
\end{center}
\end{figure}

\begin{figure}
\begin{center}
\includegraphics*[width=4in]{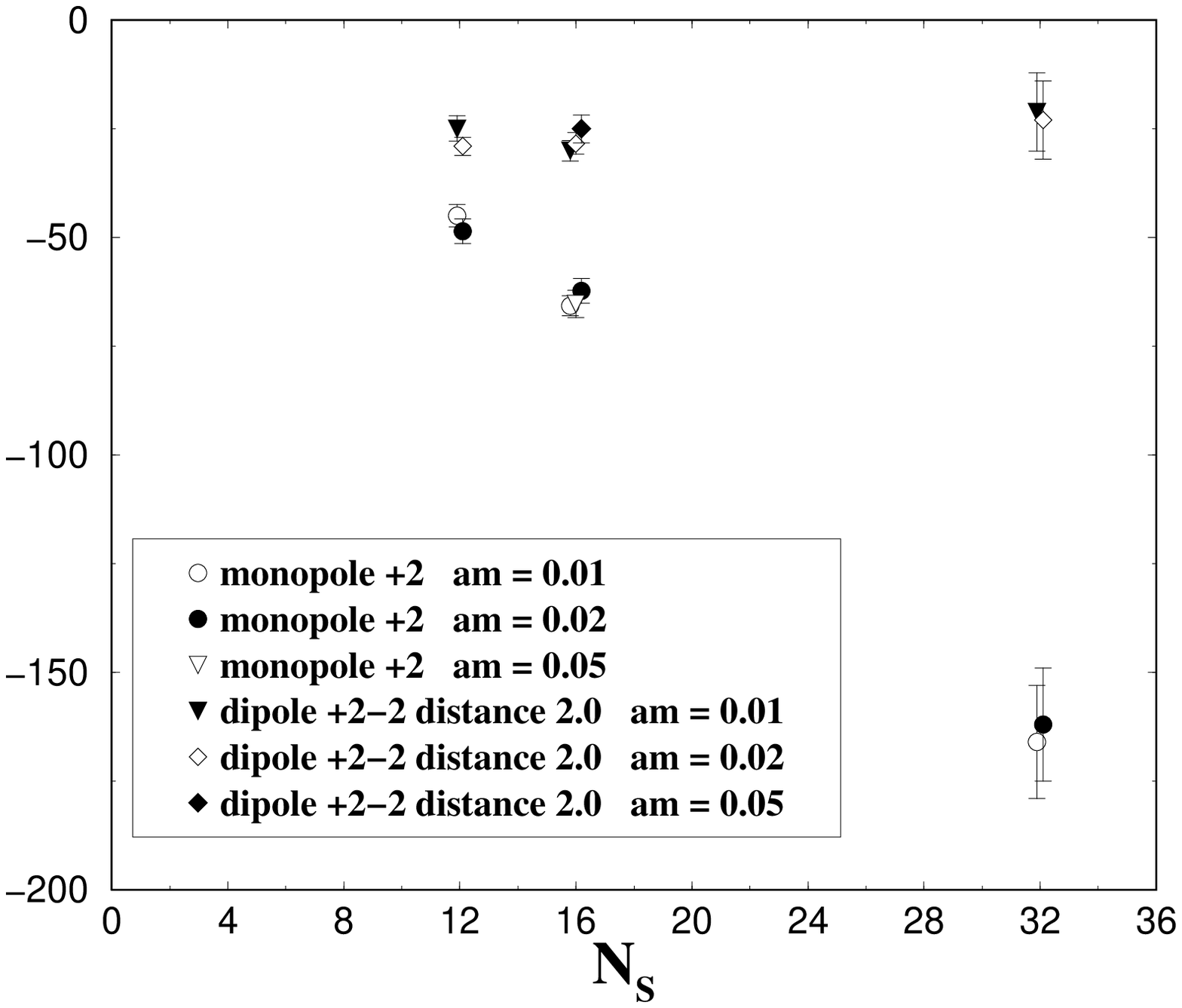}
\caption{$\rho$ tends to $-\infty$ in the thermodynamical limit, or $<\mu>\to 0$ }
\label{8}
\end{center}
\end{figure}

\begin{figure}
\begin{center}
\includegraphics*[width=4in]{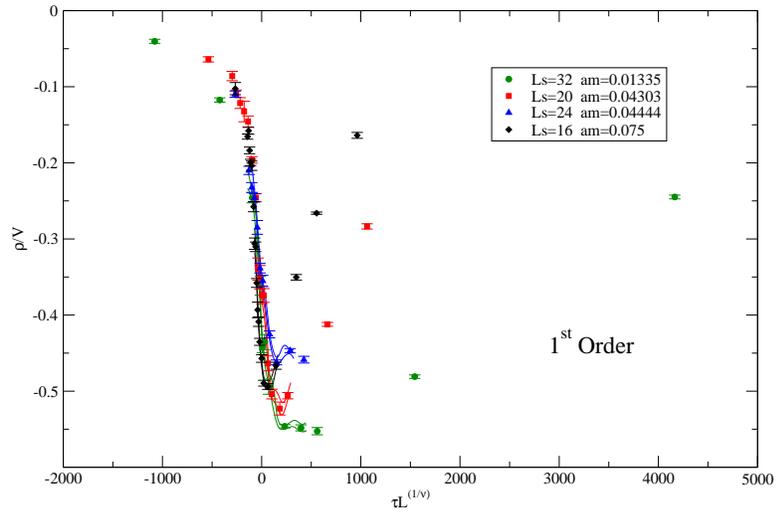}
\caption{Scaling of $\rho$ consistent with a first order transition. }
\label{9}
\end{center}
\end{figure}

   The case of $N_f=2$ $QCD$ can be approached in the same way. The results are displayed in the Fig.'s\ref{6}, \ref{7},\ref{8}, \ref{9}. The finite size scaling is that of a first order transition, and definitely excludes a second order transition in the universality class of O(4), O(2) model\cite{ddlpp}.\\
   The implications of this fact, together with a finite size scaling analysis of other quantities, like the specific heat, the chiral condensate and its susceptibility will be the object of the next section.
   
   \section{$N_f=2$  QCD}
  QCD with two flavors of light quarks is a good approximation to nature, and also a specially instructive system from the theoretical point of view. For the sake of simplicity we shall consider two quarks of equal mass $m$.\\
  The phase diagram  is shown in Fig.\ref{1}.  For    $m\ge 2Gev$  the system is quenched to all effects, the phase transition is first order and $<L>$  is a good order parameter, $Z_3$ the relevant symmetry. 
  At $m\approx 0$ a phase transition takes place from the spontaneously broken phase to a symmetric phase, and $<\bar\psi\psi>$ is the order parameter. In the intermediate region of $m$'s chiral symmetry is broken by the mass, $Z_3$ is broken by the coupling to quarks and apparently there is no order parameter.
  Also The $U_A(1)$ symmetry  which is broken by the anomaly is expected to be restored about
  at the same temperature as the chiral symmetry. Three transitions, deconfinement, chiral, $U_A(1)$:
  are they independent? Of course a definition of deconfinement is needed to answer this question.\\
  \subsection{The Chiral Transition}
  If one assumes, following reference \cite{PW} that low mass scalars and pseudoscalars are the relevant degrees of freedom, the order parameters are
  \begin{equation}   
  \underline \Phi :   \Phi_{ij}=<\bar \Psi_i(1+\gamma_5)\Psi_j>\, ,  i,j=1...N_f
  \end{equation}
  Under the symmetry group $SU(N_f)\otimes SU(N_f)\otimes U_A(1)$\\
  $$\Phi \to e^{i\alpha}U_L\Phi U_R$$.
  The most general effective Lagrangean  (density of free energy) invariant under the symmetry group is
  \begin{equation}
  L_{eff} = {1\over 2} Tr(\partial_{ \mu}\Phi^{\dagger}\partial_{\mu}\Phi) - {m^2\over 2}Tr(\Phi^{\dagger}\Phi) -{ {\pi}^2\over 3}g_1[Tr(\Phi^{\dagger}\Phi)] ^2-{ {\pi}^2\over 3}g_2[Tr(\Phi^{\dagger}\Phi)^2] +\\ c[det{\Phi} + det{\Phi}^{\dagger}]
  \end{equation}
  Terms with higher dimension have been neglected since they become irrelevant at the critical point.\\
  Infrared stable fixed points indicate second order phase transitions. A  $(4-{\epsilon})$ extrapolation to $3d$ is intended. The last term in Eq(4.2) is the Wess-Zumino term describing the anomaly: it is invariant under $SU(N_f)\otimes SU(N_f)$ but not under $U_A(1)$, and has dimension $N_f$.
  For  $N_f\geq3$  it is irrelevant and no IR stable fixed point exists, so that the transition is weak first order.\\
  For  $N_f=2$ instead the Wess-Zumino term has dimension $2$ so that its square and its product with the mass term are also relevant.  If  $c=0$ at the fixed point the symmetry is  $O(4)\otimes O(2)$
  and no IR fixed point exists, so that the transition is 1st order. Physically  this happens if the mass of the 
$\eta \prime $, $m_{\eta\prime }$,  vanishes at $T_c$. \\
  If $c\neq 0$, or if $m_{\eta\prime }$ is non zero at $T_c$, the symmetry is $O(4)$ and the transition can be second order.  If this is the case the transition is a crossover around the critical point [see Fig. \ref{1}],
  a tricritical point is expected at non zero chemical potential \cite{SS} which could be observed in heavy ion collisions. No evidence of it has emerged from experiments to date.\\
  If, instead, the transition is first order it will also be such in the vicinity of the chiral point and possibly 
  all along the transition line, and no tricritical point exists.\\
  This issue is fundamental to understand confinement: a first order phase transition is a real transition
  and can correspond to a change of symmetry and to the existence of an order parameter.\\
  A crossover means that one can go continuously from the confined region to the deconfined one and that confinement is not an absolute property of the QCD vacuum.
  \subsection{Thermodynamics}
\begin{figure}
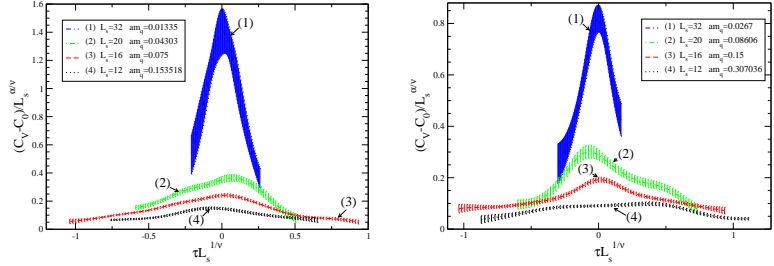

\begin{center}
\includegraphics*[width=1.9in]{f10-1.eps}\hspace{.2in}\includegraphics*[width=1.9in]{f10-2.eps}
\caption{Test of scaling Eq.(4.5), with $M= 74.7$ (left) and  $M= 149.4 $(right) : the curves should coincide.}
\label{10}
\end{center}
\end{figure}

  The order of the transition can be determined by a finite size scaling analysis \cite{fischer}\cite{brezin}
  of lattice simulations.\\
   Let $\tau = (1-{T \over T_c})$ be the reduced temperature. As $\tau \to 0$ the correlation length of the order parameter, $\xi$, diverges as \\
   $$ \xi \approx \tau^{-\nu}$$\\
   so that the ratio of the lattice spacing $a$ to $\xi$ is negligible and there is scaling. If $L_s$ is the spacial extension of the lattice, the scaling laws read
   \begin{equation} 
   C_V-C_0 \approx L_s^{\alpha\over \nu} \Phi_C(\tau L_s^{1\over\nu}, mL_s^{y_h})
   \end{equation}
   and
   \begin{equation}
   \chi \approx L_s^{\gamma \over \nu}\Phi_{\chi}(\tau L_s^{1\over\nu}, mL_s^{y_h})
   \end{equation}
   Here $C_V={{\partial E}\over {\partial T}}_V$, and $$\chi \equiv\int d^3x<\Phi(\vec x)\Phi(\vec 0)>_{conn}$$is the susceptibility of the order parameter $\Phi$.\\
   The critical indexes $\alpha, \beta, \gamma, \nu$ are the anomalous dimensions of the operators 
   and identify the order and the universality class of the transition. Eq's(4.3) and (4.4) are nothing but the 
   renormalization group equations. The subtraction needed for $C_V$ corresponds to an additive 
   renormalization\cite{brezin}.\\
   Notice that the scaling law of the specific heat is unambiguous whilst that for $\chi$ only holds if $\Phi$ is the order paramenter : the equality of the index ${\nu}$ for the two scalings can  a legitimation of the order parameter.\\
   The scaling laws Eq.'s(4.3) and (4.4) involve two scales, a fact which makes the analysis complicated with respect to the simpler case of quenched QCD.
   To simplify the problem one can study the dependence on one scale by keeping the other one fixed
   \cite {ddp}. One possibility is to vary  $m$ and $L_s$ keeping the quantity $mL_s^{1\over \nu}$
   which appears in the scaling laws fixed. The scaling equation (4.3) becomes then
   \begin{equation}
   C_V-C_0 \approx L_s^{\alpha\over \nu} \Phi_C(\tau L_s^{1\over\nu}, mL_s^{y_h}=M)
    \end{equation}
   so that the peak scales as
   \begin{equation} 
   (C_V-C_0)_{peak}\propto L_s^{\alpha\over \nu}
   \end{equation}
   This allows a determination of ${\alpha\over \nu}$. The critical index $y_h$ is the same within errors 
   for $O(4)$ and $O(2)$ universality classes, so that the same simulations can be used to check both 
   universality classes; moreover the index $\alpha$ is negative for both, implying that the peak should decrease with increasing volume. (see Table 1) 
  \begin{table}[bt!]
\begin{tabular}{|c|c|c|c|c|c|}
\hline & $y_h$ & $\nu$ & $\alpha$ & $\gamma$ & 
$\delta$\\
\hline $O(4)$ &  2.487(3) & 0.748(14) & -0.24(6) & 1.479(94) & 
4.852(24)\\
\hline $O(2)$ & 2.485(3) & 0.668(9) & -0.005(7) & 1.317(38) & 
 4.826(12)\\
\hline $MF$ & $9/4$ & $2/3$ & 0 & 1 & 3\\
\hline $1^{st} Order$ & 3 & $1/3$ & 1 & 1 & $\infty$\\
\hline
\end{tabular}
\caption{Critical exponents.}\label{CRITEXP}
\end{table}
   Fig.\ref{10} shows a test of Eq(4.5),Fig.\ref{12} a test of Eq(4.6). $O(4)$ and $O(2)$ universality classes are excluded
   with a high confidence level ($\chi^2 /dof \simeq 20$) : the peaks increase rapidly with the volume instead of decreasing.
   
For the details of the determination of the subtraction $C_0$ see  \cite{ddp}.
   \begin{figure}
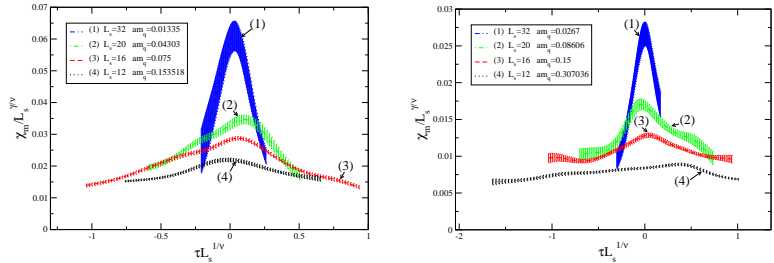

\begin{center}
\includegraphics*[width=1.9in]{f11-1.eps}\hspace{.2in}\includegraphics*[width=1.9in]{f11-2.eps}
\caption{Scaling Eq.(4.7) of the chiral susceptibility,  with $M= 74.7$ (left) and  $M= 149.4$ (right) : the curves should coincide.}
\label{11}
\end{center}
\end{figure}
   
   A similar result is obtained for the susceptibility of $<\bar\psi \psi>$ which is believed to be a good order parameter near $T_c$ (Fig.\ref{11}  and Fig. \ref{12})
   \begin{equation}
   \chi \approx \Phi_{\chi}(\tau L_s^{1\over\nu}, mL_s^{y_h}=M)
   \end{equation}
   and
   \begin{equation}
   \chi_{peak}\propto L_s^{\gamma\over \nu}
   \end{equation}
   For this test $\chi ^2/dof \approx 10$.
   \begin{figure}
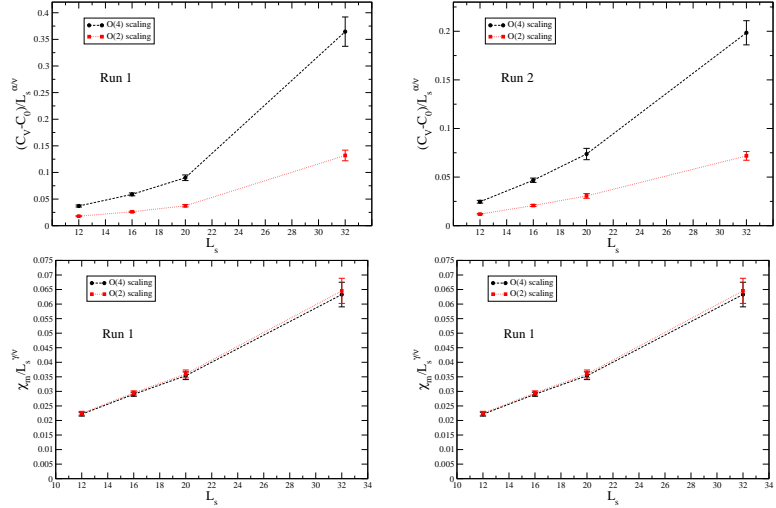

\begin{center}
\includegraphics*[width=1.9in]{f12-1.eps}\hspace{.2in}\includegraphics*[width=1.9in]{f12-2.eps}\\
\includegraphics*[width=1.9in]{f12-3.eps}\hspace{.2in}\includegraphics*[width=1.9in]{f12-3.eps}
\caption{Test of the scaling Eq.'s (4.6) and (4.8) }
\label{12}
\end{center}
\end{figure}

   As a result we can state that the transition is neither   in the universality class of $O(4)$ nor in that of $O(2)$.
    Another possibility is to look at the large volume limit keeping the first variable fixed: $\tau L_s^{1\over   \nu}$  is related to the ratio ${\xi \over L_s}$ of the correlation length to the spacial size of the lattice, while the other variable $m L_s^{y_h}$ is related to the ratio of the pion Compton wave length ${1\over {m_{\pi}}} $ to $L_s$.
    As $L_s$  goes much larger than ${1\over {m_{\pi}} }$ a finite limit is reached and\cite{ddp}
    \begin{equation}
    C_V-C_0 \approx  m^{{-\alpha}\over {\nu y_h}} f_C(\tau L_s^{1\over\nu})
    \end{equation}
    and
    \begin{equation}
     \chi \approx m^{\gamma \over {\nu y_h}}\Phi_{\chi}(\tau L_s^{1\over\nu})
     \end{equation}

\begin{figure}
\begin{center}
\includegraphics*[width=4in]{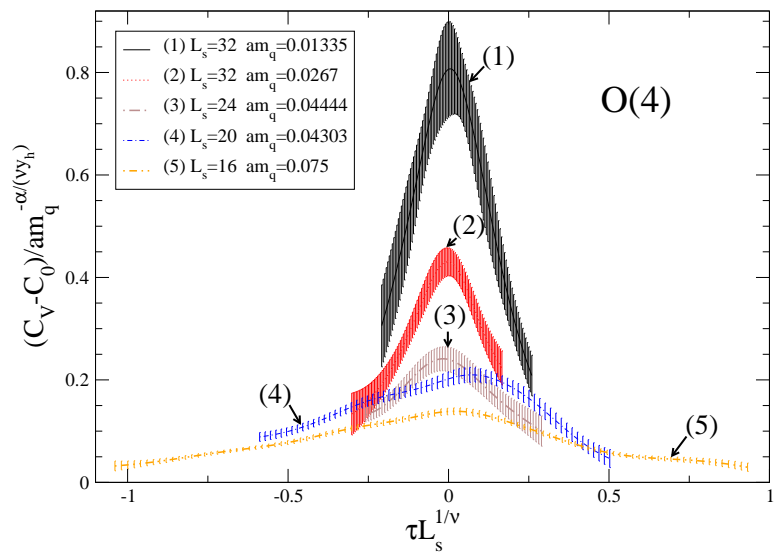}
\caption{Scaling Eq.(4.9) for O(4).}
\label{13}
\end{center}
\end{figure}

\begin{figure}
\begin{center}
\includegraphics*[width=4in]{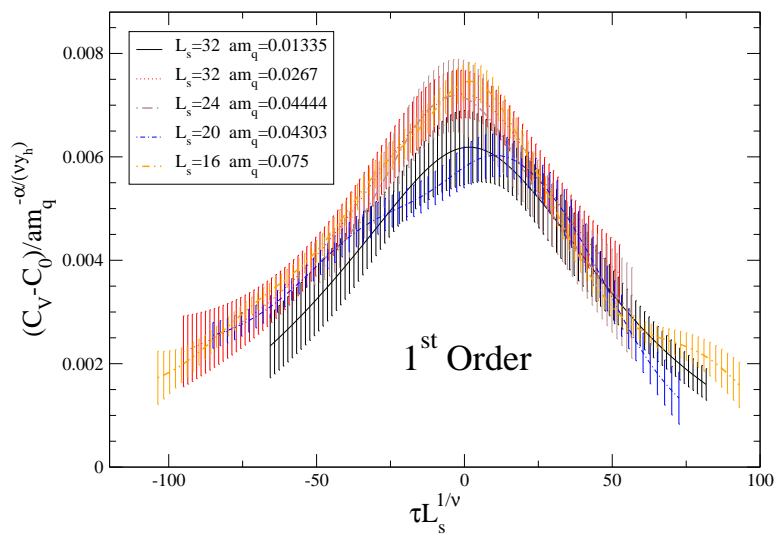}
\caption{Scaling Eq.(4.9) for first order. }
\label{14}
\end{center}
\end{figure}

     The result of this analysis is shown in Fig.\ref{13}and Fig\ref{14}. No scaling is observed assuming second order transition with $O(4)$ or $O(2)$, but a good scaling for first order.

\begin{figure}
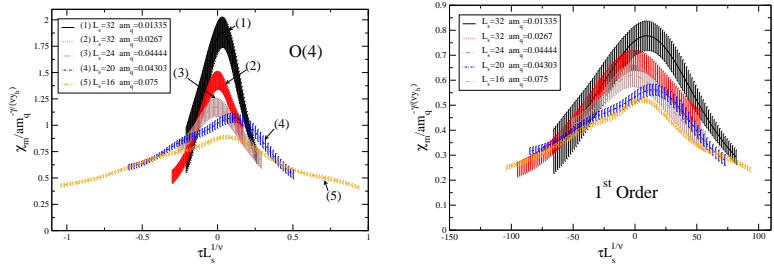

\begin{center}
\includegraphics*[width=1.9in]{f15-1.eps}\hspace{.2in}\includegraphics*[width=1.9in]{f15-2.eps}
\caption{Testing Eq.(4.10) for O(4) and first order. }
\label{15}
\end{center}
\end{figure}

 A similar result is obtained for the scaling Eq.(4.10) Fig.\ref {15}
     
\begin{figure}
\begin{center}
\includegraphics*[width=4in]{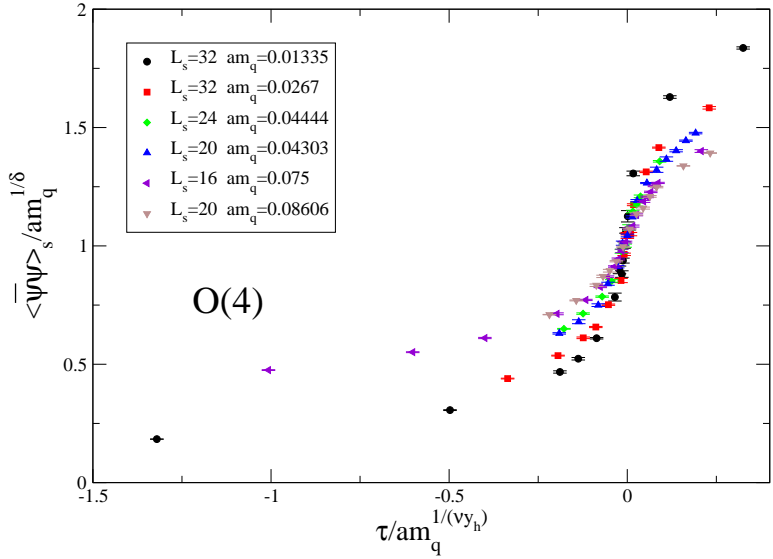}
\caption{Scaling of the magnetic equation of state (4.11) assuming O(4) }
\label{16}
\end{center}
\end{figure}

\begin{figure}
\begin{center}
\includegraphics*[width=4in]{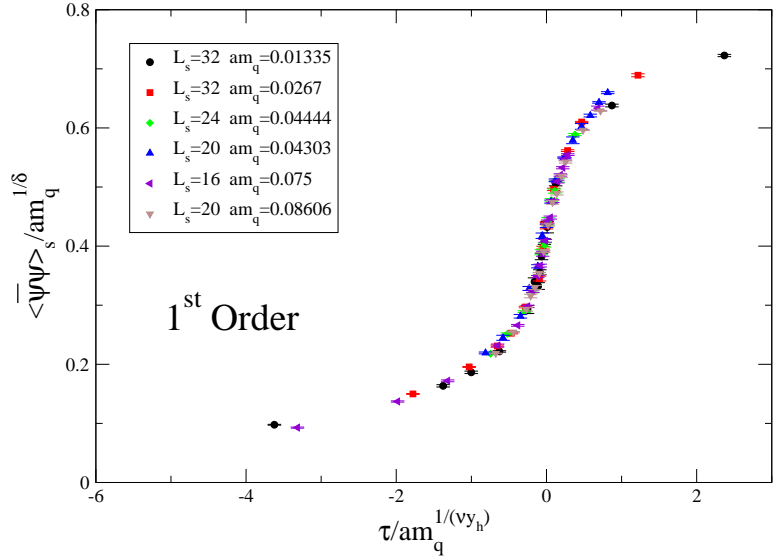}
\caption{Scaling of the magnetic equation of state (4.11) assuming first order }
\label{17}
\end{center}
\end{figure}

     Finally one can investigate the so called magnetic equation of state:
     \begin{equation}
     <\bar \psi \psi>  =  m^{1\over \delta} f(\tau m^{-1\over {\nu y_h}})
     \end{equation}
     For $O(4)$ $\delta= 4.85$, for first order $\delta = \infty$
      Again no scaling is observed assuming $O(4)$ or $O(2)$ second order transition Fig.\ref{16}, and good scaling for first order, Fig\ref{17}.

The issue is fundamental and deserves further attention.

      \section{ Concluding remarks}
      We have discussed the experimental evidence for confinement and how it naturally implies that 
      there  exists  a dual symmetry in QCD whose breaking is responsible for confinement.
       We have presented the two most accredited candidates for dual topological excitations, vortices and monopoles.\\
       We have then shown how the working hypothesis that monopoles confine via dual superconductivity of the vacuum can be tested  by numerical simulations on the lattice, through an order parameter which is the $vev $ of  operators  carrying magnetic charge. The numerical tests strongly support the validity of this idea, which can be put in a consistent form and made independent on the choice of the abelian projection. This holds both for quenched QCD and in the presence of dynamical quarks.\\
       A prerequisite is that deconfinement is a true order-disorder phase transition, and not a crossover,
       which would allow a continuous  path from confined to deconfined phase.
       We have thus discussed a test with $N_f=2$ QCD where an unsolved  dilemma exists between the existence of a crossover and a first order phase transition. We definitely exclude a second order  chiral transition which would imply a crossover at non zero quark mass, whilst we find evidence for a first order transition. The issue is fundamental and deserves further studies.\\
       From what we have seen we can conclude that the dual excitations of  $QCD$  are magnetically charged, or that dual superconductivity of the vacuum can be the mechanism of confinement.\\ However we are not yet able to identify them.

\end{document}